\documentclass[a4paper]{article}
\usepackage{amsmath}

\def\d{{\rm d}}
\newcommand{\A}{\operatorname{\cal A}}
\newcommand{\ff}{\operatorname{\mathit F}}
\newcommand{\bb}{\operatorname{\mathit B}}
\newcommand{\RR}{\operatorname{\mathit R}}

\begin{document}
\begin{titlepage}
\begin{center}
{\large\bf Autonomous multispecies reaction-diffusion systems with
more-than-two-site interactions} \vskip 2\baselineskip \centerline
{\bf Ahmad Shariati $^{a,c}$\footnote{e-mail:shahram@iasbs.ac.ir},
Amir Aghamohammadi $^{b,c}$
\footnote{e-mail:mohamadi@theory.ipm.ac.ir}, {\rm and} Mohammad
Khorrami $^{a,c}$\footnote{e-mail:mamwad@iasbs.ac.ir}}
 \vskip 2\baselineskip
{\it $^a$ Institute for Advanced Studies in Basic Sciences,}
\\ {\it P. O. Box 159, Zanjan 45195, IRAN}\\
{\it $^b$ Department of Physics, Alzahra University, Tehran 19384,
IRAN}\\
{\it $^c$ Institute for Studies in Theoretical Physics and
Mathematics,}\\
{\it P. O. Box 5531, Tehran 19395, IRAN}
\end{center}


 \vskip
1cm

\begin{abstract}
Autonomous multispecies systems with more-than-two-neighbor
interactions are studied. Conditions necessary and sufficient for
closedness of the evolution equations of the $n$-point functions
are obtained. The average number of the particles at each site for
one species and three-site interactions, and its generalization to
the more-than-three-site interactions is explicitly obtained.
Generalizations of the Glauber model in different directions,
using generalized rates, generalized number of states at each
site, and generalized number of interacting sites, are also
investigated.
\end{abstract}
\vskip 10 mm
\end{titlepage}
\section {Introduction}
The principles of equilibrium statistical mechanics are well
established. But, thermal equilibrium is a special case, and
little is known about the properties of systems not in
equilibrium, for example about the relaxation toward the
stationary state. Some interesting problems in non-equilibrium
systems are non-equilibrium phase transitions described by
phenomenological rate equations, and the way the system relaxes to
its steady state. As mean-field techniques, generally, do not give
correct results for low-dimensional systems, people are motivated
to study exactly-solvable stochastic models in low dimensions.
Moreover, solving one-dimensional systems should in principle be
easier. Exact results for some models on a one-dimensional lattice
have been obtained, for example in [1--12]. Different methods have
been used to study these models, including analytical and
asymptotic methods, mean field methods, and large-scale numerical
methods. Systems with more than one species have also been studied
[13--27]. Many of the arguments are based on simulation results.
There are, however, some exact results as well. For most of the
models studied, the interaction is between nearest neighbors.

In \cite{GS}, a 10-parameter family of stochastic models with
interactions between nearest neighbors has been studied. In these
models, the $k$-point equal time correlation functions $\langle
n_i n_j\cdots n_k\rangle $ satisfy linear differential equations
involving no higher-order correlations. We call these models
\textit{autonomous}, in the sense that the evolution equations of
$n$-point functions are closed (contain only $n$- or less-point
functions). These linear equations for the average number of the
particles $\langle n_i\rangle $ have been solved. The same models
were studied on lattices with boundaries in \cite{AM2}. It was
shown that these models may exhibit dynamic and static phase
transitions. The same idea has been generalized to multi-species
models \cite{AAMS} in one dimension with two-site interactions.
There, conditions have been obtained that the Hamiltonian should
satisfy in order that the evolution equation for correlation
functions be closed. The set of equations for average densities
can be written in terms of four matrices. These matrices are not
determined uniquely from the Hamiltonian: there is a kind of gauge
transformation one can apply on them which of course, does not
change the evolution equation. A formal solution for the average
densities of different species was found. The large-time behaviour
of the average densities of different species was also studied.
The time evolution equations for more-point functions, generally
contain not only these four matrices, but also elements of the
Hamiltonian, and to obtain a closed form for their solution is
generally not easy.

The Glauber dynamics was originally proposed to study the
relaxation of the Ising model near equilibrium states. It was also
shown that, there is a relation between the kinetic Ising model at
zero temperature and the diffusion annihilation model in one
dimension. There is an equivalence between domain walls in the
Ising model and particles in the diffusion annihilation model.
Kinetic generalizations of the Ising model, for example the
Glauber model or the Kawasaki model, are phenomenological models
and have been studied extensively [30--41].

In this paper, autonomous multispecies systems with
more-than-two-neighbor interactions are studied. Conditions
necessary and sufficient for closedness of the evolution equations
of the $n$-point functions are obtained. As an example, we obtain
explicitly the average number of the particles at each site for
one species and three-site interactions. This is then generalized
to the case where more than three site interact. As another
example, a generalization of the Glauber model is presented. In
this generalization, the processes are the same as those of the
ordinary Glauber model, but the rates depend on three free
parameters, rather than one free parameter in the ordinary Glauber
model. Finally this model is further generalized to the case where
the number of interacting sites is more than three and the number
of states at each site is more than two.

\section{Models leading to closed set of evolution equations}
The models addressed are multi-species exclusion
reaction-diffusion models. That is, each site is a vacancy or
contains one particle. There are several kinds of particles, but
at any time at most one kind can be present at each site.
Throughout the paper, the system is assumed to be translationally
invariant. Consider first a case where the interaction is between
three neighboring sites. Then the Hamiltonian describing the
system can be written as
\begin{equation}
{\cal H}=\sum_{i=1}^L H_{i,i+1,i+2}.
\end{equation}
The number of sites is $L$ and the number of possible states in a
site is $q$ (one of these states, for example the $q$-th one, may
be the vacancy); different states of each site are denoted by
$\alpha$, $\alpha =1,\cdots q$. Introducing $n^{\alpha}_i$ as the
number operator of the particles of type $\alpha$ in the site $i$,
we have
\begin{equation} \label{n1}
\sum_{\alpha =1}^q  n^\alpha_i =1.
\end{equation}
The average number of the particles of the type $\alpha$ in the
site $i$ at the time $t$ is
\begin{equation}
\langle n^{\alpha}_i \rangle = \langle  S \vert n^{\alpha}_i \vert P(t)
\rangle,
\end{equation}
where $ \vert P(t)\rangle :=\exp (t{\cal H})\ \vert P(0)\rangle $
represents the state of the system at the time $t$,
\begin{equation}
\langle S\vert =\underbrace{\langle s\vert\otimes \cdots \otimes
\langle  s\vert}_{L},
\end{equation}
and
\begin{equation}
\langle s\vert :=\underbrace{(1\ 1\ \cdots 1)}_{q}.
\end{equation}
So, the time evolution of $\langle n^\alpha_i \rangle $ is given
by
\begin{equation}\label{dn}
{\d \over \d t}\langle n^\alpha_i \rangle = \langle  S \vert
n^\alpha_i {\cal H}\vert P(t)\rangle .
\end{equation}
The only terms of the Hamiltonian ${\cal H}$ entering the above
equation are $H_{i,i+1,i+1}$, $H_{i-1,i,i+1}$, and,
$H_{i-1,i,i+1}$. The result of acting any matrix $Q$ on the bra
$\langle s\vert$ is equal to that of acting the diagonal matrix
$\tilde Q$ on the same bra, provided each diagonal element of the
matrix $\tilde Q$ is equal to the sum of all elements of the
corresponding column in the matrix $Q$. So, the actions of
$(1\otimes 1\otimes n^\alpha)H$, $(1\otimes n^\alpha\otimes 1)H$,
and $( n^\alpha\otimes 1\otimes 1)H$ on $\langle
s\vert\otimes\langle s\vert\otimes \langle s\vert$ are equal to
the actions of three diagonal matrices on $\langle
s\vert\otimes\langle s\vert\otimes\langle s\vert$. We use the
symbol $\sim$ to denote the equality of the action on $\langle
s\vert\otimes\langle s\vert\otimes\langle s\vert$. We have
\begin{align}
( n^\alpha\otimes 1\otimes 1)H &\sim \sum_{\beta\gamma\lambda}
\sideset{^1}{^\alpha_{\beta\gamma\lambda}}\A n^\beta\otimes
n^\gamma\otimes n^\lambda\nonumber\\
(1\otimes n^\alpha\otimes 1)H &\sim \sum_{\beta\gamma\lambda}
\sideset{^2}{^\alpha_{\beta\gamma\lambda}}\A n^\beta\otimes
n^\gamma\otimes n^\lambda\nonumber\\
(1\otimes 1\otimes n^\alpha)H &\sim \sum_{\beta\gamma\lambda}
\sideset{^3}{^\alpha_{\beta\gamma\lambda}}\A n^\beta\otimes
n^\gamma\otimes n^\lambda,
\end{align}
where $\sideset{^i}{^\alpha_{\beta\gamma\lambda}}\A$'s are defined
as
\begin{align}
\sideset{^1}{^\alpha_{\beta\gamma\lambda}}\A&:=s_\tau s_\omega
H^{\alpha\tau\omega}_{\beta\gamma\lambda}\nonumber\\
\sideset{^2}{^\alpha_{\beta\gamma\lambda}}\A&:=s_\tau s_\omega
H^{\tau\alpha\omega}_{\beta\gamma\lambda}\nonumber\\
\sideset{^3}{^\alpha_{\beta\gamma\lambda}}\A&:=s_\tau s_\omega
H^{\tau\omega\alpha}_{\beta\gamma\lambda}.
\end{align}
An implicit summation (from 1 to $q$) over the a same subscript
and superscript is always assumed. From these, (\ref{dn}) takes
the form
\begin{equation}\label{ndot}
\langle \dot n^\alpha_i \rangle =
\sideset{^1}{^\alpha_{\beta\gamma\lambda}}\A\langle n^\beta_i
n^\gamma_{i+1} n^\lambda_{i+2}\rangle
+\sideset{^2}{^\alpha_{\beta\gamma\lambda}}\A n^\beta_{i-1}
n^\gamma_i n^\lambda_{i+1}\rangle +
\sideset{^3}{^\alpha_{\beta\gamma\lambda}}\A\langle n^\beta_{i-2}
n^\gamma_{i-1}n^\lambda_i\rangle
\end{equation}
Generally, the right-hand side of (\ref{ndot}) contains one-,
two-, and three-point functions. (Note that $n^\alpha$'s are not
independent.) We want to obtain a condition that only one point
functions appear in the right-hand side. To do this, we consider
the expression
\begin{equation}\label{e1}
u=f_{\alpha\beta\gamma}n^\alpha_i n^\beta_j n^\gamma_k,
\end{equation}
and ask for the condition that the right-hand side is expressible
in terms of linear combinations of $n$'s, provided
\begin{equation}\label{e2}
s_\alpha n^\alpha_l=1.
\end{equation}
It is obvious that if
\begin{equation}\label{e3}
f_{\alpha\beta\gamma}=\sideset{_1}{_\alpha}\ff +
\sideset{_2}{_\beta}\ff +\sideset{_3}{_\gamma}\ff ,
\end{equation}
then the right hand side of (\ref{e1}) is expressible in terms of
linear combinations of $n$'s. To prove that this form for $f$ is
necessary as well, we just count the number of independent
variables in $f$'s satisfying the desired property. One can write
$n^q_l$ in terms one 1 and other $n^\alpha_l$'s. Then it is seen
that a general cubic form of $n$'s is expressible in terms of
$q^3$ independent forms of $n$'s, each containing no more than
three $n$'s. Of these, $1+3(q-1)$ expressions (the monomials of
degree zero and one of $3(q-1)$ independent variables) are
desirable. The coefficients of other monomials should be zero. So,
from $q^3$ independent variables in $f$, there remains only
$3(q-1)+1$ independent variables in $f$'s satisfying the desired
condition. It seems that the right-hand side of (\ref{e3})
contains more independent variables, $3q$. But we note that the
transformation
\begin{equation}\label{e4}
\sideset{_i}{_\alpha}\ff\to\sideset{_i}{_\alpha}\ff +
\sideset{_i}{}\bb,
\end{equation}
does not change the right-hand side of (\ref{e3}), provided
\begin{equation}\label{e5}
\sum_i\sideset{_i}{}\bb=0.
\end{equation}
This means that there are $3-1$ redundant variables in the
expression (\ref{e3}). So (\ref{e3}) contains actually the correct
number of independent variables, and hence is the most general
form of $f$ with the desired property.

So, in order that (\ref{ndot}) be expressible in terms of only
one-point functions, one must have
\begin{equation}\label{sbc}
\sideset{^i}{^\alpha_{\beta\gamma\lambda}}\A=
\sideset{^i_1}{^\alpha_\beta}\A+\sideset{^i_2}{^\alpha_\gamma}\A
+\sideset{^i_3}{^\alpha_\lambda}\A .
\end{equation}
It is noted that $\sideset{^i_j}{}\A$'s are not determined
uniquely. Applying the \textit{gauge} transformation
\begin{equation}\label{e6}
\sideset{^i_j}{^\alpha_\beta}\A\to\sideset{^i_j}{^\alpha_\beta}\A +
\sideset{^i_j}{^\alpha}\bb =\sideset{^i_j}{^\alpha_\beta}\A +
\sideset{^i_j}{^\alpha}\bb s_\beta,\qquad\hbox{with}\qquad
\sum_j \sideset{^i_j}{^\alpha}\bb =0
\end{equation}
does not change the right-hand side of (\ref{sbc}).

If (\ref{sbc}) is satisfied, (\ref{dn}) takes the form
\begin{align}\label{ndot2}
\langle \dot n^\alpha_i \rangle =&(\sideset{^1_1}{^\alpha_\beta}\A +
\sideset{^2_2}{^\alpha_\beta}\A +\sideset{^3_3}{^\alpha_\beta}\A )
\langle n^\beta_i\rangle\nonumber\\
&+(\sideset{^1_2}{^\alpha_\beta}\A +\sideset{^2_3}{^\alpha_\beta}\A)
\langle n^\beta_{i+1}\rangle +(\sideset{^2_1}{^\alpha_\beta}\A +
\sideset{^3_2}{^\alpha_\beta}\A)\langle n^\beta_{i-1}\rangle\nonumber\\
&+\sideset{^1_3}{^\alpha_\beta}\A\langle n^\beta_{i+2}\rangle
+\sideset{^3_1}{^\alpha_\beta}\A\langle n^\beta_{i-2}\rangle .
\end{align}
(\ref{sbc}) in fact guarantees that the time-evolution equations
of $n$-point functions contain only $n$- and less-point functions.
In the simplest case, the one-species, each site is vacant or
occupied by only one kind of particles. Then, the matrices
$\sideset{^i_j}{}\A$ are two by two.

One can do the same arguments for the case where more than three
neighboring sites interact. Suppose the number of interacting
sites is $N$. One defines
\begin{equation}\label{e7}
\sideset{^i}{^{\alpha_i}_{\beta_1\cdots\beta_N}}\A :=
\left(\prod_{l\ne i}s_{\alpha_l}\right)
H^{\alpha_1\cdots\alpha_N}_{\beta_1\cdots\beta_N}.
\end{equation}
To ensure that in the time-evolution equation of one-point
functions only one-point functions appear, one must have
\begin{equation}\label{e8}
\sideset{^i}{^{\alpha_i}_{\beta_1\cdots\beta_N}}\A =
\sum_j\sideset{^i_j}{^{\alpha_i}_{\beta_j}}\A .
\end{equation}
Here too, the \textit{gauge} transformation (\ref{e6}) does not
change the right-hand side of (\ref{e8}), and hence the physics of
the problem. It is to be noted that (\ref{e8}) is also sufficient
for $n$-point functions evolution equations to contain no more
than $n$-point functions.
\subsection{Some special cases}
We now consider some special cases.
\subsubsection{single species case}
In this case the matrices $\sideset{^i_j}{}\A$ are two by two. The
time-evolution equation for $\langle  n_k\rangle$ will then be
\begin{equation}
\langle\dot n_i\rangle =-\alpha\langle n_i\rangle +\beta\langle
n_{i+1}\rangle+\beta'\langle n_{i-1}\rangle +\gamma\langle n_{i+2}
\rangle +\gamma'\langle n_{i-2}\rangle +\delta .
\end{equation}
$\delta$ can be eliminated using the redefinition
\begin{equation}\label{nm}
x_i:=\langle n_i\rangle -
{\delta\over{\beta +\beta'+\gamma +\gamma'-\alpha}}.
\end{equation}
Then, introducing the generating function
\begin{equation}
G(z,t)=\sum_{-\infty}^{\infty}x_i z^i,
\end{equation}
one arrives at,
\begin{equation}
\dot G(z,t)=\left( -\alpha +\beta z^{-1}+\beta' z+\gamma z^{-2}
+\gamma' z^2\right) G(z,t),
\end{equation}
the solution to which is
\begin{equation}
G(z,t)=\exp\left[ t\left( -\alpha +\beta z^{-1}+\beta' z +\gamma
z^{-2}+\gamma' z^2\right)\right] G(z,0).
\end{equation}
Using
\begin{equation}
\exp\left[{u\over 2}\left( z+z^{-1}\right)\right] =
\sum_{k=-\infty}^\infty {\rm I}_k(u)z^k,
\end{equation}
one arrives at
\begin{equation}
x_k(t)=e^{-\alpha t}\sum_{j,l=-\infty}^{\infty} {\rm I}_{k-j-2l}
\left(2t\sqrt{\beta\beta'}\right){\rm I}_l
\left(2t\sqrt{\gamma\gamma'}\right) (\beta'/\beta)^{(k-j-2l)/2}
(\gamma'/\gamma)^{l/2} x_j(0).
\end{equation}
A similar procedure can be done for
more-than-three-neighboring-sites interactions. The main
difference will be the number of modified Bessel functions
appearing in the expression.
\subsubsection{Generalizations of the Glauber model}
Consider a two-state three-neighbor interaction of the form
\begin{equation}
\sideset{^1}{^\alpha_{\beta\gamma\lambda}}\A =
\sideset{^3}{^\alpha_{\beta\gamma\lambda}}\A =0.
\end{equation}
This means that, similar to the Glauber model, any site interacts
only with its neighboring sites. The interactions are
\begin{align}
AAA\longrightarrow & A\emptyset A\quad &\mu_1\nonumber\\
\emptyset\emptyset\emptyset\longrightarrow &
\emptyset A\emptyset\quad &\mu_2\nonumber\\
A\emptyset A\longrightarrow & AAA\quad &\lambda_1\nonumber\\
\emptyset A\emptyset\longrightarrow &\emptyset\emptyset\emptyset\quad
&\lambda_2\nonumber\\
AA\emptyset\longrightarrow & A\emptyset\emptyset\quad &
\alpha_1\nonumber\\
\emptyset\emptyset A\longrightarrow &\emptyset AA\quad &
\alpha_2\nonumber\\
A\emptyset\emptyset\longrightarrow & AA\emptyset\quad &
\beta_1\nonumber\\
\emptyset AA\longrightarrow &\emptyset\emptyset A\quad &\beta_2.
\end{align}
This is a generalization of the Glauber model. For the ordinary
Glauber model
\begin{align}
\mu_1 &=\mu_2=1-\tanh\frac{J}{k_{\rm B}T}\nonumber\\
\lambda_1 &=\lambda_2=1+\tanh\frac{J}{k_{\rm B}T}\nonumber\\
\alpha_1 &=\alpha_2=\beta_1=\beta_2=1
\end{align}
The criterion (\ref{e8}) for the closedness of the
evolution-equation for one-point functions results in the
following relations between the rates in the generalized Glauber
model.
\begin{align}
\mu_i -\alpha_i&=\beta_j -\lambda_j\quad\hbox{for any $i,j$}
\nonumber\\
\alpha_1+\beta_1&=\alpha_2+\beta_2
\end{align}
So there are four independent variables in terms of them the above
eight parameters can be expressed. One can write the expressions
as
\begin{align}\label{e25}
\mu_1&=A-B-C-C'\nonumber\\ \mu_2&=A+B\nonumber\\
\lambda_1&=A+B+C+C'\nonumber\\ \lambda_2&=A-B\nonumber\\
\alpha_1&=A-B-C'\nonumber\\ \alpha_2&=A+B+C\nonumber\\
\beta_1&=A+B+C'\nonumber\\ \beta_2&=A-B-C.
\end{align}
Of course one of the parameters can be absorbed through a
time-rescaling. The resulting evolution-equation for the average
particle-number is
\begin{equation}\label{e26}
\langle\dot n_i\rangle =A+B -2A\langle n_i\rangle +C\langle
n_{i+1}\rangle +C'\langle n_{i-1}\rangle ,
\end{equation}
which is easy to solve.

This generalized Glauber model can further be generalized in two
directions: when the number of the interacting sites is more than
3, and when the number of states of each site is more than 2. The
first case means that the interaction is in a block of length $N$,
resulting in the change of the state of a single specific site in
that block. This rate of change depends on the states of this site
and the states of the other $N-1$ sites. Let's label this specific
site of the block by $0$. (Usually the length of the block is
considered to be an odd integer ($2k+1$), and the evolving site is
assumed to be the central one.) Denote the state of the site $i$
by $\sigma_i$, where $\sigma_i$ can take the values $1$ (particle)
or $0$ (vacancy). Then the evolution equation for the average
particle number is
\begin{align}\label{hasan}
\langle \dot{n}_0\rangle =&-\left\langle\sum_{\vec{\sigma}}
R(1,\vec{\sigma})\, n_0 \prod_{i\neq 0} \left[ 1-n_i+\sigma_i(2
n_i -1)\right]\right\rangle\nonumber\\
&+\left\langle\sum_{\vec{\sigma}}R(0, \vec{\sigma})(1-n_0)
\prod_{i\neq 0}\left[ 1-n_i+\sigma_i(2 n_i
-1)\right]\right\rangle.
\end{align}
Here the state of other interacting sites is denoted by
$\vec{\sigma}$, $R(\sigma_0, \vec{\sigma})$ is the rate of change
of the state of site $0$, from $\sigma_0$ to $1-\sigma_0$, when
the states of the other interacting sites is $\vec{\sigma}$.

We are looking for those rates $R(\sigma_0, \vec{\sigma})$ that
make the right-hand side of this evolution equation a linear
combination of $\langle n_i\rangle$'s. The claim is that the
general form of these rates is
\begin{equation}\label{e30}
R(\sigma_0, \vec{\sigma})=A+(-1)^{\sigma_0}\left( B+\sum_{i \neq 0}
C_i\, \sigma_i\right) .
\end{equation}
Inserting this ansatz into the evolution equation (\ref{hasan}) we
get
\begin{equation}
\langle\dot{n}_0\rangle =B+A\langle 1-2 n_0\rangle + \sum_{i\neq
0}C_i\langle n_i\rangle .
\end{equation}
So it is clear that the ansatz (\ref{e30}) leads into a closed set
of evolution-equations for the average particle number. It remains
to prove that this ansatz is the most general one satisfying this
property. To see this, one considers (\ref{hasan}). In the
right-hand side of this equation there are $2^N$ terms (the
expectation of monomials in terms of $n_i$'s). Of these, we wish
that the coefficients of all be zero, except for the coefficients
of the constant term, and linear terms. So, there are $2^N-(N+1)$
equations to be satisfied for the rates (consisting themselves of
$2^N$ unknowns). This shows that the rates satisfying the desired
property contain $N+1$ independent variables, and it is clear that
the ansatz (\ref{e30}) contains $N+1$ independent variables. So it
is the most general solution.

Now consider the second generalization, when the number of
possible states at each site is more than 2, say $q$, and each
block consists of $N$ sites. The state of the site $i$ is denoted
by $\sigma_i$, which can take $q$ values. That site the state of
which is denoted by $0$. The rate of change of the state of he
site $0$ from $\beta$ to $\alpha$ is denoted by
$R^\alpha_{\beta,\vec{\sigma}}$. The evolution equation for the
average numbers is then
\begin{align}
\langle\dot{n}^\alpha_0\rangle =&\left\langle\sum_{\vec{\sigma}}
\sum_{\beta\neq\alpha}R^\alpha_{\beta,\vec{\sigma}}\, n_0^\beta
\prod_{i\ne 0}n_i^{\sigma_i}\right\rangle\nonumber\\
&-\left\langle\sum_{\vec{\sigma}}\sum_{\beta\neq\alpha}
R^\beta_{\alpha,\vec{\sigma}}\, n_0^\alpha \prod_{i\ne
0}n_i^{\sigma_i}\right\rangle .
\end{align}
Defining
\begin{equation}\label{hosein}
R^\alpha_{\alpha,\vec{\sigma}}:=-\sum_{\beta\neq\alpha}
R^\beta_{\alpha,\vec{\sigma}},
\end{equation}
the evolution equation reads
\begin{equation}
\langle\dot{n}^\alpha_0\rangle =\sum_{\vec{\sigma}}
R^\alpha_{\beta,\vec{\sigma}}\left\langle n_0^\beta
\prod_{i\ne 0}n_i^{\sigma_i}\right\rangle .
\end{equation}
It is easy to see that for the right-hand side of this equation be
expressible in terms of one-point functions, one should have
\begin{equation}
R^\alpha_{\beta,\vec{\sigma}}=\sideset{_0}{^\alpha_\beta}\RR +
\sum_{i\ne 0}\sideset{_i}{^\alpha_{\sigma_i}}\RR
\end{equation}
Note that this is nothing but
\begin{equation}
\sideset{^0}{^\alpha_{\beta,\vec{\sigma}}}\A =
\sum_{i}\sideset{^0_i}{^\alpha_{\sigma_i}}\A ,
\end{equation}
obtained in (\ref{e8}). In this case, other $\sideset{^i}{}\A$'s
vanish. The evolution equation of the one-point functions is then
\begin{equation}
\langle\dot n^\alpha_0\rangle =
\sum_i\sideset{_i}{^\alpha_\beta}\RR\langle n^\beta_i\rangle .
\end{equation}
Previous discussions shows that this is the most general form of
the rates for them the evolution-equations of the one-point
functions contain only one-point functions. Here too, there is a
\textit{gauge} freedom in choosing $\sideset{_i}{}\RR$'s, namely
\begin{equation}
\sideset{_i}{^\alpha_\beta}\RR\to \sideset{_i}{^\alpha_\beta}\RR +
\sideset{_i}{^\alpha}\bb s_\beta,\qquad\hbox{with}\qquad \sum_i
\sideset{_i}{^\alpha}\bb =0.
\end{equation}
This is the same \textit{gauge} freedom encountered with earlier,
(\ref{e6}).



\newpage
\begin{thebibliography}{99}
\bibitem{ADHR} F. C. Alcaraz, M. Droz, M. Henkel, \& V. Rittenberg;
               Ann. Phys. {\bf 230} (1994) 250.
\bibitem{KPWH} K. Krebs, M. P. Pfannmuller, B. Wehefritz, \&
               H. Hinrichsen; J. Stat. Phys. {\bf 78}[FS] (1995) 1429.
\bibitem{HS1}  H. Simon; J. Phys. {\bf A28} (1995) 6585.
\bibitem{PCG}  V. Privman, A. M. R. Cadilhe, \& M. L. Glasser; J. Stat.
               Phys. {\bf 81} (1995) 881.
\bibitem{HOS1} M. Henkel, E. Orlandini, \& G. M. Sch\"utz; J. Phys.
               {\bf A28} (1995) 6335.
\bibitem{HOS2} M. Henkel, E. Orlandini, \& J. Santos; Ann. of Phys.
               {\bf 259} (1997) 163.
\bibitem{AL}   A. A. Lusknikov; Sov. Phys. JETP {\bf 64} (1986) 811.
\bibitem{AKK}  M. Alimohammadi, V. Karimipour, \& M. Khorrami; Phys. Rev.
               {\bf E57} (1998) 179.
\bibitem{RK}   F. Roshani \& M. Khorrami; Phys. Rev. {\bf E60} (1999) 3393.
\bibitem{RK2}  F. Roshani \& M. Khorrami; cond-mat/0007352.
\bibitem{AKK2} M. Alimohammadi, V. Karimipour, \& M. Khorrami; J. Stat.
               Phys. {\bf 97} (1999) 373.
\bibitem{AM1}  A. Aghamohammadi \& M. Khorrami; J. Phys. {\bf A33} (2000) 7843.
\bibitem{SBH}  R. Schoonover,D. Ben--Avraham,S. Havlin, R. Kopelman, \&
               G. Weiss;  Physica {\bf A171} (1991) 232.
\bibitem{TW}   D. Toussaint, F. Wilczek; J. Chem. Phys. {\bf 78} (1983) 2642.
\bibitem{KR}   K. Kong, S. Render; Phys. Rev. Lett. {\bf 52} (1984) 955.
\bibitem{BPZ}  D. Ben-Avraham, V. Privman, D. Zhong; Phys. Rev. {\bf E52}
               (1995) 6889.
\bibitem{IKR}  I. Ispolatov, P. L. Krapivsky, S. Render; Phys. Rev. {\bf E52}
               (1995) 25406889.
\bibitem{LP}   J. W. Lee, V. Privman; J. Phys. {\bf A30} (1997) L317.
\bibitem{ZK}   Z. Koza; cond-mat/9601072.
\bibitem{VK}   V. Karimipour; Phys. Rev. {\bf E59} (1999) 205.
\bibitem{KC}   K. Kim, K. H. Chang; J. Phys. Soc. Jpn. {\bf 68} (1999) 1450.
\bibitem{AAMS} A. Aghamohammadi, A. H. Fatollahi, M. Khorrami, \& A.
               Shariati; Phys. Rev. {\bf E62} (2000) 4642.
\bibitem{KK1}  M. Khorrami \& V. Karimipour; J. Stat. Phys. {\bf 100} (2000) 999.
\bibitem{OM}   G. Odor, N. Menyhard; Phys. Rev. E. 61 (2000) 6404.
\bibitem{AA}   M. Alimohammadi, N. Ahmadi; Phys. Rev. {\bf E62} (2000) 1674.
\bibitem{FJ}   F. H. Jafarpour; cond-mat/0004357.
\bibitem{RK3}  F. Roshani \& M. Khorrami; cond-mat/0101127.
\bibitem{GS}   G. M. Sch\"utz; J. Stat. Phys. {\bf 79}(1995) 243.
\bibitem{AM2}  A. Aghamohammadi \& M. Khorrami; cond-mat/0012495.
\bibitem{RG}   R. J. Glauber; J. Math. Phys. {\bf 4} (1963) 294.
\bibitem{KK}   K. Kawasaki; Phys. Rev. {\bf 145} (1966) 224.
\bibitem{AF}   J. G. Amar \& F. Family; Phys. Rev. {\bf A41} (1990) 3258.
\bibitem{TV}   T. Vojta; Phys. Rev {\bf E55} (1997) 5157.
\bibitem{SSG}  R. B. Stinchcombe, J. E. Santos, \& M. D. Grynberg; J. Phys.
               {\bf A31} (1998) 541.
\bibitem{GL}   C. Godr\`eche \& J. M. Luck; J. Phys. {\bf A33} (2000) 1151.
\bibitem{MA}   M. Khorrami \& A. Aghamohammadi; Phys. Rev. {\bf E63} (2001) 042102.
\bibitem{GO}   L. L. Gon\c{c}alves \& A. L. Stella; J. Phys.
               {\bf A20} (1987) L387.
\bibitem{DRS}  M. Droz, Z. R\'acz, \& J. Schmidt; Phys. Rev. {\bf A39}
               (1989) 2141.
\bibitem{GF}   B. C. S. Grandi \& W. Figueiredo; Phys. Rev. {\bf E53}
               (1996) 5484.
\bibitem{AT}   S. Artz \& S. Trimper; Int. J. Mod. Phys. {\bf B12} (1998) 2385.
\bibitem{MA2}  M. Khorrami \& A. Aghamohammadi; cond-mat/0103090.
\end{thebibliography}
\end{document}